\documentclass[aps,preprint,showpacs,showkeys,nofootinbib,amsfonts,amssymb,amsmath]{revtex4}

\newcommand{\blambda}{\mbox{\boldmath $\lambda$}}

\begin{document}

\preprint{Yukawa Institute Kyoto} \preprint{YITP-08-85}

\title{Shape invariance in prepotential approach\\ to exactly solvable models}

\author{Choon-Lin Ho
\footnote{email:~hcl@mail.tku.edu.tw\\
Tel:~+886-2-26215656\\
Fax:~+886-2-2620-9917}
}
\affiliation{ Yukawa Institute for Theoretical Physics, Kyoto
University, Kyoto 606-8502, Japan\\and\\
Department of Physics, Tamkang University, Tamsui 251, Taiwan,
Republic of China\footnote{Permanent address}}

\date{Jan 9, 2009}

\begin{abstract}
In supersymmetric quantum mechanics, exact-solvability of
one-dimensional quantum systems can be classified only with an
additional assumption of integrability, the so-called shape
invariance condition. In this paper we show that in the
prepotential approach we proposed previously, shape invariance is
automatically satisfied and needs not be assumed.
\end{abstract}

\pacs{03.65.Ca, 03.65.Ge, 02.30.Ik}
\keywords{Prepotential, exact solvability, shape invariance,
supersymmetry}

\maketitle

\section{Introduction}

It is generally known that exactly solvable systems are very rare
in any branch of physics. Thus any new method to construct exactly
solvable models would be of interest to the community concerned.
It is therefore very interesting to realize that most exactly
solvable one-dimensional quantum systems can be obtained in the
framework of supersymmetric quantum mechanics (SUSYQM)
\cite{Cooper,Junker}. However, in SUSYQM, exact-solvability can be
classified only with an additional assumption of integrability, so
called shape invariance (SI) condition \cite{Gend}.  Hence in
SUSYQM the SI condition must be taken as a sufficient condition
for integrability at the outset. What is more, the transformation
of the original coordinate, say $x$, to a new one $z=z(x)$ needed
in solving the SI condition is not naturally determined within the
framework of SUSYQM in most cases, but have to be taken as given
from the known solutions of the respective models. It would be
more satisfactory if the exact-solvability of a quantal system,
including the required change of coordinates, could be determined
with the simplest, and the most natural requirements.

In \cite{Ho1,Ho2,Ho3} a unified approach to both the exactly and
quasi-exactly solvable systems is presented. This is a simple
constructive approach, based on the so-called prepotential
\cite{CS,CS1,CS2,CS3,CS4,CS5,Ho4,Ho5,Ho6}, which gives the
potential as well as the eigenfunctions and eigenvalues
simultaneously.  The novel feature of the approach is that both
exact and quasi-exact solvabilities can be solely classified by
two integers, the degrees of two polynomials which determine the
change of variable and the zero-th order prepotential.  Hence this
approach treats both quasi-exact and exact solvabilities on the
same footing, and it provides a simple way to determine the
required change of coordinates $z(x)$. All the well-known exactly
solvable models given in \cite{Cooper,Junker}, most quasi-exactly
solvable models discussed in \cite{TU,Tur,GKO,Ush,Ush1}, and some
new quasi-exactly solvable ones (also for non-Hermitian
Hamiltonians), can be generated by appropriately choosing the two
polynomials.

Since all the well-known one-dimensional exactly solvable models
obtained in SUSYQM, by taking SI condition as a sufficient
condition, can also be derived without the SI condition in the
prepotential approach, one wonders what role the SI condition
plays in the latter approach.  In this paper we would like to show
that the SI condition is only a necessary condition in the
prepotential approach to exactly solvable systems.  Therefore,
unlike SUSUQM, shape invariance needs not be assumed in the
prepotential approach..

This paper is organized as follows. In Sect.~II we give a brief
review of the prepotential approach to exactly solvable models
with both sinusoidal and non-sinusoidal coordinates. The idea of
SI as a sufficient condition of integrability in SUSYQM is
sketched in Sect.~III. Sect.~IV and V then demonstrate that in the
prepotential approach for models with sinusoidal and
non-sinusoidal coordinates, SI is automatically satisfied and
needs not be imposed. Sect.~VI concludes the paper.

\section{Prepotential approach}

The main ideas of the prepotential approach can be summarized as
follows (we adopt the unit system in which $\hbar$ and the mass
$m$ of the particle are such that $\hbar=2m=1$). Consider a wave
function $\phi_N(x)$ ($N$: non-negative integer) which is defined
as
\begin{eqnarray}
\phi_N(x)\equiv e^{-W_0(x)}p_N (z),\label{phi_N}
\end{eqnarray}
with
\begin{eqnarray}
p_N (z)\equiv \left\{
               \begin{array}{ll}
                1, & N=0;\\
                \prod_{k=1}^N (z-z_k),& N>0.
                        \end{array}
               \right.
\label{phi}
\end{eqnarray}
Here $z=z(x)$ is some real function of the basic variable $x$,
$W_0(x)$ is a regular function of $z(x)$, and $z_k$'s are the
roots of $p_N(z)$. The variable $x$ is defined on the full line,
half-line, or finite interval, as dictated by the choice of
$z(x)$. The function $p_N(z)$ is a polynomial in an
$(N+1)$-dimensional Hilbert space with the basis $\langle
1,z,z^2,\ldots,z^N \rangle$. $W_0(x)$ defines the ground state
wave function.

The wave function $\phi_N$ can be recast as
\begin{eqnarray}
 \phi_N =\exp\left(- W_N(x,\{z_k\})
\right), \label{f2}
\end{eqnarray}
with $W_N$ given by
\begin{eqnarray}
W_N(x,\{z_k\}) = W_0(x) - \sum_{k=1}^N \ln |z(x)-z_k|. \label{W}
\end{eqnarray}
Operating on $\phi_N$ by the operator $-d^2/dx^2$ results in a
Schr\"odinger equation $H_N\phi_N=0$, where
\begin{eqnarray}
H_N &=&-\frac{d^2}{dx^2} + V_N,\\
V_N&\equiv&  W_N^{\prime 2} - W_N^{\prime\prime}.
\end{eqnarray}
Here prime represents differentiation with respect to $x$. It is
seen that the potential $V_N$ is defined by $W_N$, and we shall
call $W_N$ the $N$th order prepotential.  From Eq.~(\ref{W}), one
finds that $V_N$ has the form $V_N=V_0+\Delta V_N$:
\begin{eqnarray}
V_0 &=&W_0^{\prime 2} - W_0^{\prime\prime},\nonumber\\
 \Delta V_N &=&
-2\left(W_0^\prime z^\prime
-\frac{z^{\prime\prime}}{2}\right)\sum_{k=1}^N \frac{1}{z-z_k} +
\sum_{{k,l}\atop{k\neq l}} \frac{z^{\prime 2}}{(z-z_k)(z-z_l)}.
\label{V}
\end{eqnarray}

Thus the form of $V_N$, and consequently its solvability, are
determined by the choice of $W_0(x)$ and $z^{\prime 2}$ (or
equivalently by $z^{\prime\prime}=(dz^{\prime 2}/dz)/2$). Let
$W_0^\prime z^\prime=P_m(z)$ and $z^{\prime 2}=Q_n(z)$ be
polynomials of degree $m$ and $n$ in $z$, respectively. In
\cite{Ho1}, it was shown that if the degree of $W_0^\prime
z^\prime$ is no higher than one ($m\leq 1$), and the degree of
$z^{\prime 2}$ no higher than two ($n\leq 2$), then in $V_N(x)$
the parameter $N$ and the roots $z_k$'s, which satisfy the
so-called Bethe ansatz equations (BAE) to make the potential
analytic, will only appear in an additive constant and not in any
term involving powers of $z$. Such system is then exactly
solvable. If the degree of one of the two polynomials exceeds the
corresponding upper limit, the resulted system is quasi-exactly
solvable.  The transformed coordinates $z(x)$ such that the degree
of $z^{\prime 2}$ is no higher than two are called sinusoidal
coordinates. There are six types of one-dimensional exactly
solvable models which are based on such coordinates, namely, the
shifted-oscillator, three-dimensional oscillator, Morse, Scarf
type I and II, and generalized P\"oschl-Teller models as list in
\cite{Cooper}.

In \cite{Ho3}, the prepotential approach to exactly solvable
systems was extended to systems based on non-sinusoidal
transformed variable $z(x)$ which is a solution of $z^\prime
=\lambda - z^2$.  With this, the remaining four types of exactly
solvable systems listed in \cite{Cooper}, namely, the Coulomb,
Eckart, and Rosen-Morse type I and II models, are also covered by
the prepotential approach.

\subsection{Sinusoidal coordinates}

For exactly solvable models with sinusoidal coordinates we take
$m=1$ and $n=2$, i.e., $P_1(z)=az+b$, and $Q_2(z)=\alpha z^2
+\beta z +\gamma$, where $a,~b,~\alpha,~\beta$ and $\gamma$ are
real constants. With these choices we obtain \cite{Ho1}
\begin{eqnarray}
 V_N ={W_0^\prime}^2 - W_0^{\prime\prime}+ \alpha N^2 -2aN -2\sum_{k=1}^N
\frac{1}{z-z_k}\left\{\left(a -\frac{\alpha}{2}\right)z_k +b-
\frac{\beta}{4} - \sum_{l\neq k} \frac{Q_2(z_k)}{z_k-z_l}\right\}.
\label{dV1}
\end{eqnarray}
Demanding the residues at $z_k$'s vanish gives the set of Bethe
ansatz equations
\begin{eqnarray}
\left(a -\frac{\alpha}{2}\right)z_k +b- \frac{\beta}{4} -
\sum_{l\neq k} \frac{Q_2(z_k)}{z_k-z_l}=0,~~k=1,2,\ldots,N.
\label{BAE}
\end{eqnarray}
With this set of roots $z_k$, the last term in Eq.~(\ref{dV1})
vanishes, and we obtain a potential $V_N(x)=V_0(x)-E_N$ without
simple poles.  Here $V_0(x)=W_0^{\prime 2}-W_0^{\prime\prime}$
does not involve $N$ and $z_k$'s, and can be taken as the exactly
solvable potential of the system with eigen-energies
$E_N=2aN-\alpha N^2$. In fact, $V_0(x)$ is exactly the
supersymmetric form presented in \cite{Cooper} for the
shifted-oscillator, three-dimensional oscillator, Morse, Scarf
type I and II, and generalized P\"oschl-Teller models (for easy
comparison, we note that $\alpha$ and $a$ here equal $\pm\alpha^2$
and $\alpha A$ in \cite{Cooper}).

\subsection{Non-sinusoidal coordinates}

As mentioned before, the Coulomb, Eckart, and Rosen-Morse type I
and II models involve a change of coordinates of the form
$z^\prime=\lambda-z^2$ which is non-sinusoidal.  But with a slight
extension of the methods in \cite{Ho1}, all these four models can
be treated in a unified way in the prepotential approach
\cite{Ho3}. The extension is simply to allow the coefficients in
$W_0$ be dependent on $N$. It turns out that $W_0^\prime$ takes
the form
\begin{eqnarray}
W_0^\prime (N)=-\left(A+N\alpha\right) z + \frac{B}{A+N\alpha},
\label{W0}
\end{eqnarray}
where $A$ and $B$ are real parameters. Then the potential $V_N$
becomes $V_N(x)=V(x)-E_N$, where
\begin{eqnarray}
V(x)=A\left(A-1\right)z^2(x) -2B z(x),\label{V0}
\end{eqnarray}
and
\begin{eqnarray}
 E_N =
 -\frac{B^2}{\left(A+N\right)^2}-\lambda\left[A\left(2N+1\right)
 +N^2\right].\label{E_N}
\end{eqnarray}
Now $V(x)$ is independent of $N$, and can be taken to be the
potential of an exactly solvable system, with eigenvalues $E_N$
($N=0,1,2,\ldots$).  The corresponding wave functions $\phi_N$ are
given by (\ref{phi_N}):
\begin{eqnarray}
\phi_N\sim e^{\left(A+N\right)\int^x dx
z(x)-\frac{B}{A+N}x}~p_N(x),~~N=0,1,\ldots \label{phi_N2}
\end{eqnarray}
The BAE satisfied by the roots $z_k$'s are
\begin{eqnarray}
\sum_{l\neq k}\frac{z_k^2-\lambda}{z_k-z_l} -\left(A+N-1\right)z_k
+\frac{B}{A+N} =0, ~~k=1,2,\dots,N. \label{BAE2}
\end{eqnarray}

Finally,we mention here that $V(x)$ in (\ref{V0}) can be obtained,
up to an additive constant, from $W_0(N)$ with any value of $N$.
Particularly, the form adopted in supersymmetric quantum mechanics
(e.g., in \cite{Cooper}) is obtained from the zero-th order
prepotential $W_0(N=0)$ with $N=0$ \cite{Ho3}.

\section{Shape invariance in supersymmetric quantum mechanics}

From the discussions in the last section, we see that in the
prepotential approach, exactly solvable models are determined by
the zero-th order prepotential $W_0(x)$ in the sinusoidal cases,
or $W_0\equiv W_0(N=0)$ with $N=0$ in the four non-sinusoidal
cases. The potential $V_0$ is completely determined by $W_0$:
$V_0={W_0^\prime}^2 - W_0^{\prime\prime}$, and consequently, the
Hamiltonian $H_0=-d^2/dx^2+V_0$ is factorizable  as $H_0=A^{+}A$
with the first-order operators
\begin{eqnarray}
A\equiv\frac{d}{d x}+W^\prime_0,\qquad
 A^{+}\equiv -\frac{d}{d x}+W^\prime_0.
\label{A_N}
\end{eqnarray}
This fact is indeed the base of SUSYQM.  In SUSYQM
\cite{Cooper,Junker} one considers the relation between the
spectrum of $H_0$ and that of its so-called super-partner
Hamiltonian $H_1$ constructed according to $H_1\equiv
AA^+=-d^2/dx^2+ V_1$, where $V_1\equiv W_0^{\prime ^2} +
W_0^{\prime\prime}$.  In forming $V_1$, it is equivalent to using
a prepotential $-W_0$. The ground state of $H_1$ is therefore
$\exp(W_0)$, and it follows that the ground states of $H_0$ and
$H_1$ cannot be both normalizable.

Let us suppose the ground state of $H_0$, i.e. $\exp(-W_0)$, is
normalizable, and denote the normalized eigenfunctions of the
Hamiltonians $H_{0,1}$ by $\psi_{n}^{(0,1)}$ with eigenvalues
$E_{n}^{(0,1)}$, respectively. Here the subscript $n=0,1,2,\ldots$
denotes the number of nodes of the wave function. It is easily
proved that $V_0$ and $V_1$ have the same energy spectrum except
for the ground state of $V_0$ with $E_0^{(0)}=0$, which has no
corresponding level for $V_1$ \cite{Cooper,Junker}. More
explicitly, we have the following supersymmetric relations:
\begin{gather}
E_{n}^{(1)}=E_{n+1}^{(0)},\nonumber\\
\psi_{n}^{(1)}=\bigl(E_{n+1}^{(0)}\bigr)^{-1/2}
 A\psi_{n+1}^{(0)},~~A\psi_{0}^{(0)}=0,\label{SUSY}\\
\psi_{n+1}^{(0)}=\bigl(E_{n}^{(1)}\bigr)^{-1/2}
 A^{+}\psi_{n}^{(1)}.\nonumber
\end{gather}
Hence $A$ annihilates $\psi_{0}^{(0)}$, and converts an
eigenfunction of an excited state of $H_0$ into an eigenfunction
of $H_1$ with the same energy, but with one less number of nodes,
while $A^{+}$ does the reverse.  Consequently, if the spectrum of
one system is exactly known, so is the spectrum of the other.

This is, however, all that supersymmetry says about the two
partner potentials. If any one of the spectra is unknown, then
supersymmetry is useless in solving them. It is therefore
gratifying that most of the well-known one-dimensional exactly
solvable models process a property called shape invariance.  With
hindsight, one can then impose shape invariance as an additional
requirement along with supersymmetry to classify exactly solvable
systems having such property.  This has been done and most exactly
solvable systems are then unified within the framework of SUSYQM
\cite{Cooper,Junker}.

Shape invariance means that the two super-partner potentials $V_0$
and $V_1$ are related by the relation
\begin{eqnarray}
V_1(x;\blambda_0)=V_0(x;\blambda_1)+R(\blambda_0), \label{shape}
\end{eqnarray}
where $\blambda_{0}$ is a set of parameters of the original $V_0$,
$\blambda_1=f(\blambda_0)$ is a function of $\blambda_0$, and
$R(\blambda_0)$ is a constant which depends only $\blambda_0$.
This implies
\begin{eqnarray}
W_0^{\prime 2}(x,\blambda_0) +
W_0^{\prime\prime}(x,\blambda_0)=W_0^{\prime 2}(x,\blambda_1) -
W_0^{\prime\prime}(x,\blambda_1) +R(\blambda_0).\label{SI-Cond}
\end{eqnarray}
Eq.~(\ref{shape}) implies that $V_1$ has the same shape as that of
$V_0$, but is defined by parameters $\blambda_1$ instead of
$\blambda_0$.  From (\ref{SI-Cond}) one deduces that the ground
state wave function of $V_1$ is $\psi_0^{(1)}\sim
\exp(-W_0(x,\blambda_1)$ with energy $R_0(\blambda_0)$. Then from
(\ref{SUSY}) we know the energy of the first excited state of
$V_0$ to be $R(\blambda_0)$, and the wave function
$\psi_1^{(0)}\sim A^{+}\psi_0^{(1)}$.  By repeated use of the
shape invariance condition, one can construct the partner $V_2$ of
$V_1$, $V_3$ of $V_2$, etc. The ground state wave function of
$V_n$ ($n=0,1,\ldots$) is $\psi_0^{(n)}\sim
\exp(-W_0(x,\blambda_n)$, where $\blambda_n=f^n(\blambda_0)$, with
energy $\sum_{k=0}^{n-1} R(\blambda_k)$.  Then again from
(\ref{SUSY}) we know that the wave function of the $n^{th}$ state
of $H_0$ is $\psi_n^{(0)}\sim (A^{+})^n\psi_0^{(n)}$, with energy
\begin{equation}
E_n^{(0)}=\sum_{k=0}^{n-1} R(\blambda_k),
~~n=0,1,\ldots\label{E-SUSY}
\end{equation}
So with shape invariance one obtains the complete spectrum of
$H_0$.

It is now obvious that SI is a sufficient condition of
integrability in SUSYQM. To classify shape-invariant exactly
solvable models in SUSYQM, one must solve the SI condition
(\ref{SI-Cond}) to get all the functional forms of $W_0(x)$,
$\blambda_1=f(\blambda_0)$, and $R(\blambda_0)$.  This general
problem is very difficult and, to the best of our knowledge, is
still unsolved. Further constraints on the possible class of shape
invariant potentials are required. Particularly, in order to
obtain the well-known exactly solvable models one must assume that
(again with hindsight) the parameters of the two partner
potentials are related by simply a translational shift, i.e.
$\blambda_1=f(\blambda_0)=\blambda_0+ {\bf m}$ differ from
$\blambda_0$ only by a set of constants ${\bf m}$. Even with this
simplification, the required change of coordinates $z=z(x)$ needed
in solving the SI condition cannot be determined naturally in the
approach of SUSYQM, but has to be taken as given from the known
solutions of the respective models.

On the other hand, in the prepotential approach SI needs not be
imposed, and $W_0$ and $z(x)$ are determined by simply picking two
polynomials with the appropriate degrees. In this sense it appears
to us that the prepotential approach is conceptually much simpler.
Nevertheless, putting the differences of the two approaches aside,
one could not help but wonder what role SI plays in the
prepotential approach. Below we would like to demonstrate that for
the exactly solvable models obtained in the prepotential approach,
SI is automatically satisfied. We shall discuss the cases with
sinusoidal and non-sinusoidal coordinates separately.

\section{Shape invariance in Prepotential approach: Sinusoidal coordinates}

Our strategy is to show that, with $z(x)$ and $W_0(x)$ given in
Sect.~II(A) and (B) that produce the ten well-known exactly
solvable models, the SI condition (\ref{SI-Cond}) is always
satisfied, i.e. one can always find the set of new parameters
$\blambda_1$ in terms of the old ones $\blambda_0$. In the
process, we demonstrate that the change in the parameters of the
shape-invariant potentials are translational.

In this section, we first consider the cases involving sinusoidal
coordinates. For exactly solvable systems, we must take
$W_0^\prime z^\prime=P_1(z)$.  Labelling the corresponding
parameters of the two shape-invariant potentials by $k=0,1$, we
have
\begin{gather}
z^{\prime 2}=Q_2(z)=\alpha z^2 +\beta z + \gamma;\\
P^{(k)}_1(z)=a_k z+b_k,,~~k=0,1\\
W^\prime_0(\blambda_k)=\frac{P^{(k)}_1(z)}{\sqrt{Q_2(z)}},~~\blambda_k=(a_k,b_k).
\end{gather}
Note that $z(x)$ is the same for the shape-invariant potentials.
Then the SI condition (\ref{SI-Cond}) leads to
\begin{eqnarray}
\left(P^{(0)2}_1-P^{(1)2}_1\right)+Q_2
\frac{d}{dz}\left(P^{(0)}_1+P^{(1)}_1\right)
-\frac{1}{2}\frac{dQ_2}{dz}\left(P^{(0)}_1+P^{(1)}_1\right)=R(\blambda_0)Q_2.
\end{eqnarray}
Equating the coefficients of the powers of $z$, one arrives at the
following equations relating the parameters
\begin{gather}
a_0^2-a_1^2=R\alpha,\nonumber\\
2\left(a_0b_0-a_1b_1\right)+\frac{\beta}{2}\left(a_0+a_1\right)
-\alpha\left(b_0+b_1\right)=R\beta,\label{SI-Cond-1}\\
b_0^2-b_1^2+\gamma\left(a_0+a_1\right)-\frac{\beta}{2}
\left(b_0+b_1\right)=R\gamma.\nonumber
\end{gather}
For simplicity we write $R$ for $R(\blambda_0)$. We mention here
that the signs of $a$ and $b$ are fixed by the normalization of
the wave functions. This means they are the same for the two
shape-invariant partner potentials.

We would like to solve (\ref{SI-Cond-1}) for
$\blambda_1=(a_1,b_1)$ and $R$ in terms of $\blambda_0=(a_0,b_0)$.
To facilitate solution, we find it convenient to first determine
all inequivalent types of sinusoidal coordinates.

\subsection{Inequivalent sinusoidal coordinates}

Depending on the presence of the parameters $\alpha,~\beta$ and
$\gamma$, there are three inequivalent cases of sinusoidal
coordinates: (i) $z^{\prime 2}=\gamma\neq 0$, (ii) $z^{\prime
2}=\beta z + \gamma$ ($\beta\neq 0$), and (iii) $z^{\prime
2}=\alpha z^2 + \beta z + \gamma$ ($\alpha\neq 0$).  By an
appropriate shifting and/or scaling, these cases can be recast
into three canonical forms.

The form given for case (i) is already the canonical form of this
case.  We shall take $\gamma> 0$ as $\gamma\leq 0$ leads to
physically uninteresting change of variable. This case gives rise
to the shifted oscillator.

By shifting $z$ to $\hat{z}\equiv z+\gamma/\beta$ in case (ii), we
get the canonical form ${\hat z}^{\prime 2}=\beta {\hat z}$.  For
physical systems we require $\beta>0$. This case corresponds to
the three-dimensional oscillator.

Case (iii) can be recast as ${\tilde z}^{\prime 2}=\alpha {\tilde
z}^2 + \tilde{\gamma}$, where $\tilde{z}\equiv z+\beta/2\alpha$
and $\tilde{\gamma}\equiv \Delta/4\alpha$ with the discriminant
$\Delta\equiv 4\alpha\gamma-\beta^2$. For the case $\Delta=0$ (the
exponential case) and $\alpha>0$, the system thus generated is
related to the Morse potential.  For $\Delta\neq 0$, we have two
situations. If $\alpha>0$ (the hyperbolic case), the canonical
form is ${\hat z}^{\prime 2}=\alpha(\hat{z}^2\pm 1)$, where
$\hat{z}\equiv \sqrt{4\alpha^2/|\Delta|} {\tilde z}$, and the plus
(minus) sign corresponds to $\Delta>0$ ($\Delta<0$). The plus sign
gives rise to the Scarf II model, while the minus sign corresponds
to the generalized P\"oschl-Teller model.  For $\alpha<0$ (the
trigonometric case), the canonical form is ${\hat z}^{\prime
2}=|\alpha|(\pm 1 -\hat{z}^2)$, where again $\hat{z}\equiv
\sqrt{4\alpha^2/|\Delta|} {\tilde z}$, and the plus (minus) sign
corresponding to $\Delta<0$ ($\Delta>0$). With the plus sign we
get the Scarf I model, while the minus sign does not lead to any
viable system as the transformation is imaginary.

From the above discussions, we see that we need only to discuss
the three inequivalent canonical cases, namely, (i) $z^{\prime
2}=\gamma\neq 0$, (ii) $z^{\prime 2}=\beta z$ ($\beta> 0$), and
(iii) $z^{\prime 2}=\alpha (z^2  + \delta)$ ($\delta=0,\pm 1$ for
$\alpha>0$, and $\delta=-1$ if $\alpha<0$).

\subsection{Case (i): $z^{\prime
2}=\gamma>0$}

For this case, it is easy to check that $a_0$ ($a_1$) must not
vanish, or it will lead to vanishing potential.  Furthermore, we
must have $a_0>0$ and $a_1>0$ in order that the wave functions be
normalizable.  The SI conditions (\ref{SI-Cond-1}) become
\begin{gather}
(a_0+a_1)(a_0-a_1)=0,\label{i-1}\\
a_0b_0-a_1b_1=0,\label{i-2}\\
b_0^2-b_1^2+\gamma\left(a_0+a_1\right)=R\gamma.\label{i-3}
\end{gather}
Equations (\ref{i-1}) and (\ref{i-2}) require $a_1= a_0,~b_1=b_0$,
or $a_1= -a_0,~b_1=-b_0$. In the latter solution the signs of
$a_1$ and $b_1$ are different from those of $a_0$ and $b_0$, and
hence the wave functions of one of the two systems cannot be
normalizable if those of the other system can.  In fact, for this
case we have $R=0$ from (\ref{i-3}). This means the ground states
of the two systems have the same energy. But the flip of both
signs of $a$ and $b$ of $W_0$ means that the ground states of the
two systems have the forms $\exp(-W_o)$ and $\exp(+W_0)$.  They
cannot be both normalizable. This is exactly the result in SUSYQM.

So we are left with the choice $a_1=a_0,~b_1=b_0$.  From
(\ref{i-3}) we have $R=2a_0$.  Thus $R$ is a constant, and from
(\ref{E-SUSY}) it implies oscillator-like spectrum, i.e.
$E_n=na_0$. This gives the shifted oscillator.

The above discussion shows that in this case SI is a necessary
condition. The parameters of the two partner systems are related
by $(a_1,~b_1)=(a_0,~b_0)$, and the shift parameter is $R=2a_0$.

\subsection{Case (ii): $z^{\prime 2}=\beta z$ ($\beta> 0$)}

Normalizability of wave functions in this case require that $a>0$
and $b<0$. Now the SI conditions (\ref{SI-Cond-1}) are
\begin{gather}
(a_0+a_1)(a_0-a_1)=0,\label{ii-1}\\
2\left(a_0b_0-a_1b_1\right)+\frac{\beta}{2}\left(a_0+a_1\right)=R\beta,\label{ii-2}\\
\left(b_0+b_1\right)\left(b_0-b_1-\frac{\beta}{2}\right) =0.
\label{ii-3}
\end{gather}
Possible solutions of these equations are $a_0\pm a_1=0$,
$b_0+b_1=0$ or $b_0-b_1-\beta/2=0$.  To keep the signs of $a$ and
$b$ unchanged, we can only take $(a_1,~b_1)=(a_0,~b_0-\beta/2)$ as
the viable solution. Then from (\ref{ii-3}) we get $R=2a_0$, which
again gives an oscillator-like spectrum.  This is just the case of
the three-dimensional oscillator.

\subsection{Case (iii): $z^{\prime 2}=\alpha (z^2  + \delta)$}

Next we consider the case with $z^{\prime 2}=\alpha (z^2  +
\delta)$ ($\delta=0,\pm 1$ for $\alpha>0$, and $\delta=-1$ if
$\alpha<0$).  As mentioned before, this case covers the Morse,
generalized P\"oschl-Teller, and the Scarf I and II potentials.
The SI conditions (\ref{SI-Cond-1}) are
\begin{gather}
a_0^2-a_1^2=R\alpha,\label{iii-1}\\
2\left(a_0b_0-a_1b_1\right)
-\alpha\left(b_0+b_1\right)=0,\label{iii-2}\\
b_0^2-b_1^2+\alpha\delta\left(a_0+a_1\right)=R\alpha\delta.
\label{iii-3}
\end{gather}

To solve $a_1~,b_1$ and $R$ in terms of $a_0$ and $b_0$, we
eliminate $R\alpha$ in (\ref{iii-3}) using (\ref{iii-1}) to get
\begin{eqnarray}
\left(b_0+b_1\right)\left(b_0-b_1\right)
+\delta\left(a_0+a_1\right)\left(a_1-a_0+\alpha\right)=0.
\label{iii-4}
\end{eqnarray}

From (\ref{iii-4}) we can have four possible sets of solutions:
\begin{eqnarray}
a_0+a_1=0,~~b_0+b_1=0;\\
a_0+a_1=0,~~b_0-b_1=0;\\
a_0-a_1=\alpha,~~b_0+b_1=0;\\
a_0-a_1=\alpha,~~b_0-b_1=0.
\end{eqnarray}
The first three sets of solutions involve change of signs of $a$
and/or $b$, and so are not viable as discussed before.  Thus for
this case we must take $(a_1,~b_1)=(a_0-\alpha,~b_0)$  which also
satisfies (\ref{iii-2}).  Eq.~(\ref{iii-1}) then gives
\begin{eqnarray}
R(\blambda_0)=\frac{a_0^2-a_1^2}{\alpha}=2a_0-\alpha.\label{R-iii}
\end{eqnarray}
From (\ref{E-SUSY}) the energies are
\begin{eqnarray}
E_n&=&\frac{a_0^2-a_n^2}{\alpha}\nonumber\\
&=&\frac{a_0^2-\left(a_0-n\alpha\right)^2}{\alpha}, ~~n=0,1,\ldots
\end{eqnarray}
This is exactly the results in SUSYQM \cite{Cooper}.

To conclude this section, we have shown that SI is automatically
satisfied in the prepotential approach for the sinusoidal cases.

\section{Shape invariance in Prepotential approach: Non-sinusoidal coordinates}

In this case, $W_0^\prime=-Az+B/A$ and
$z^\prime=\alpha(\lambda-z^2)$. Here $\blambda_0=(A,B)$. As in the
last section, we show that one can always find a set of new
parameter $\blambda_1=(A^\prime,B^\prime)$ in terms of
$\blambda_0$ that solves the SI condition (\ref{SI-Cond}). In
fact, from  (\ref{SI-Cond}) one finds
\begin{eqnarray}
A\left(A+\alpha\right)&=&A^\prime\left(A^\prime-\alpha\right),\label{SI-2-1}\\
B&=&B^\prime,\label{SI-2-2}\\
\frac{B^2}{A^2}-\alpha\lambda A &=& \frac{B^{\prime 2}}{A^{\prime
2}}+\alpha\lambda A^\prime +R.\label{SI-2-3}
\end{eqnarray}
Solutions of (\ref{SI-2-1}) are $A^\prime =-A$ and
$A^\prime=A+\alpha$.  The first solution has the sign of $A$
changed, and will lead to non-normalized wave functions.  Hence
the viable solution is $\blambda_1=(A^\prime,B^\prime)=(A+\alpha,
B)$.  Once again, the change in the parameters $A$ and $B$ of the
shape-invariant potentials are translational.  Finally, from
(\ref{SI-2-3}) we find
\begin{eqnarray}
R(\blambda_0)=B^2\left[\frac{1}{A^2}-\frac{1}{(A+\alpha)^2}\right]-\alpha\lambda\left(2A
+\alpha\right).
\end{eqnarray}
This agrees with the results in SUSYQM \cite{Cooper}.

Thus we have shown that in the prepotential approach for models
based on non-sinusoidal coordinates, SI is also a necessary
consequence of the forms of $W_0$ and $z^\prime$.

\section{Summary}

A unified approach to both the exactly and quasi-exactly solvable
systems has been proposed previously based on the so-called
prepotential in \cite{Ho1,Ho2,Ho3}. In this approach solvability
of a quantal system can be solely classified by two integers, the
degrees of two polynomials which determine the change of variable
and the zero-th order prepotential.  All the well-known exactly
solvable models obtained in SUSYQM can be easily constructed by
appropriately choosing the two polynomials.

But all these exactly solvable models are obtained in SUSYQM only
by taking the SI condition as a sufficient condition.  The
requirement to get exactly solvable models in the prepotential
approach appears to be much simpler, and definitely without the
need of SI condition.  In this paper we have shown that the SI
condition is in fact only a necessary condition in the
prepotential approach to exactly solvable systems, and hence needs
not be assumed.  In the process, we have demonstrated that the
change in the parameters of the well-known shape-invariant
potentials are indeed translational, a result which was also
assumed in SUSYQM.

\begin{acknowledgments}

This work is supported in part by the National Science Council
(NSC) of the Republic of China under Grant Nos. NSC
96-2112-M-032-007-MY3 and NSC 95-2911-M-032-001-MY2. Part of the
work was done during my visit to the Yukawa Institute for
Theoretical Physics (YITP) at the Kyoto University supported under
NSC Grant No. 97-2918-I-032-002.  I would like to thank R. Sasaki
and the staff and members of YITP for their hospitality.  I am
also grateful to Y. Hosotani and Y. Matsuo for useful discussion
and hospitality.

\end{acknowledgments}

\end{document}